%% file: COVID19_Analysis_Pakistan_SIR.tex
\documentclass[a4paper, 12pt]{article}

\usepackage[a4paper, total={7in, 9.6in}]{geometry}
\usepackage{stackengine}
\usepackage{amsmath}
\usepackage{graphicx}
\usepackage[T1]{fontenc}
\usepackage[utf8]{inputenc}
\usepackage{authblk}
\usepackage{float}
\usepackage{epstopdf}
\usepackage{caption}
\usepackage{float}
\usepackage{relsize}
\usepackage{graphicx}
\usepackage{float}
\usepackage{epstopdf}
\usepackage[margin=1cm]{caption}
\usepackage{subcaption}
\usepackage{csquotes}
\usepackage[usenames, dvipsnames]{color}
\usepackage{color}
\usepackage[pdftex, bookmarks, colorlinks,pdfstartview=FitH]{hyperref}
\usepackage{cite}
\usepackage{amssymb}
\usepackage{amsthm}
\usepackage{amsmath}
\usepackage{enumerate}
\usepackage{placeins}
\usepackage{xcolor}
\usepackage[T1]{fontenc} 
\usepackage{color}
\usepackage{array}
\usepackage{longtable}
\usepackage{authblk}
\providecommand{\keywords}[1]{\textbf{\textit{Index terms---}} #1}

\begin{document}

\title{\Large{Analysis and Prediction of COVID-19 Pandemic in  Pakistan using Time-dependent SIR  Model}}


\author[1,2]{\small{Muhammad Waqas}}
\author[1]{\small{Muhammad Farooq}}
\author[3]{\small{Rashid Ahmad}}
\author[1]{\small{Ashfaq Ahmad  \thanks{Corresponding author: Ashfaq.Ahmad@ncp.edu.pk}} }

\affil[1]{\small{Exp. High Energy Physics Department (EHEPD), National Centre for Physics \protect \\ Islamabad, Pakistan}}
\affil[2]{\small{High Energy Physics Group, Pakistan Institute of Nuclear Science and Technology \protect\\
Nilore, Islamabad, Pakistan}}
\affil[3]{\small{Department of Physics, Kohat University of Science and Technology, Kohat, Pakistan}}


\date{}

\maketitle

\abstract{
The current outbreak is known as Coronavirus Disease or COVID-19  caused by the virus SAR-COV-2 which continues to wreak havoc across the globe. The World Health Organization (WHO) has declared the outbreak a pandemic. In Pakistan, the spread of the virus is on the rise with the number of infected people and causalities rapidly increasing. In the absence of proper vaccination and treatment, to reduce the number of infections and casualties, the only option so far is to educate people regarding preventive measures and to enforce countrywide lock-down. Any strategy about the preventive measures needs to be based upon detailed analysis of the COVID-19 outbreak and accurate scientific predictions. In this paper, we conduct mathematical and numerical analysis to come up with reliable and accurate predictions of the outbreak in Pakistan. The time-dependent Susceptible-Infected-Recovered (SIR)  model is used to fit the data and provide future predictions. The turning point of the peak of the pandemic is defined as the day when the transmission rate becomes less than the recovering rate. We have predicted that the outbreak will reach its maximum peak occurring from late May to 9 June with unrecovered number of Infectives in the range 20000-47000 and the cumulative number of infected cases in the range of 57500-153100. The number of Infectives will remain at the lower end in the lock-down scenario but can rapidly double or triple if the spread of the epidemic is not curtailed and localized.  The uncertainty on single day projection in our analysis after April 15 is found to be within 5\%.
The reproduction number R$_{0}$ becomes less than unity after the peak or turning point of the outbreak. After the peak date, the infection rate will start decreasing but it might take months for the epidemic to completely fade away with 97\% recovery happening in late August-to-September 2020.
}

\vspace{1.5in}

\newpage
\section{Introduction}
\label{sec:intro}
The Corona pandemic that began in the city of Wuhan in South China in early December 2019 has now become the global pandemic. The cause of the virus outbreak was later on identified as a novel coronavirus known  as SAR-COV-2. The WHO has already declared the outbreak as a Public Health Emergency of International Concern and subsequently as a pandemic.
In the last three months, the pandemic has spread around the world and has infected around three and half million people with about 250,000 causalities worldwide.
The infected cases of COVID-19 are rapidly increasing in Pakistan as well and the positive cases have been reported from all parts of the country.
Out of the total number of 19103 infections, as recorded on May 2, 4817 have recovered whereas 440 individuals have died.

\par The rising numbers of infections can overburden the already fragile health care system of the country in the coming months if the spread is not controlled.
Any future treatment or vaccination of the COVID-19 will likely take at least a year to be available to the public. In the meantime, the only way to curtail the spread of COVID-19 is through precautionary measures and countrywide lock-downs. Such measures bring economic problems with themselves and are not easy to implement without economic losses. Therefore, an informed and effective decisions by policymakers based on a proper modeling of the pandemic can reduce the spread of the infection.
\par In the last few months studies have been carried out to understand the spread of the disease.  For example, the Susceptible-Exposed-Infectious-Removed (SEIR) model is used to model the outbreak in the city of Wuhan China \cite{Lina}. Analyses about Pakistan have been conducted but lack the reasonable estimates of the pandemic. 
The simplest version of the SIR model seems to be working better than other variants, in particular, the prediction made for China in \cite{LZhong} was found to be in agreement with the real data and hence used as a starting point. We followed a similar strategy to implement the time-dependent SIR model. 
Besides the fitting method in \cite{LZhong}, we checked the machine learning approach as documented in \cite{YChen} but that method does not give superior  results.
The SIR model depends upon two tunable  parameters $\beta(t)$, that measures the transmissions per unit time and  $\gamma(t)$ that measures the recoveries per unit time. The proper estimation of these parameters can predict a reasonable number of infections and the peak of the pandemic.  The exponential fit is used to estimate their values from data and then predicting future trends.

\par This manuscript is organized in the following way. In section \ref{sec:model}, the mathematical model used for prediction is outlined. In  section \ref{sec:dataAna}, the data used for the study in this paper and the analysis methods are discussed. Validation of the analysis machinery is discussed in section~\ref{sec:anaValid}. The results are presented in section \ref{sec:results}. The main conclusions are presented in the last section \ref{sec:conclusion}.

\section{The Time Dependent Susceptible-Infected-Recovered (SIR) Model}
\label{sec:model}
In the standard SIR model, the total population is divided  into three groups. The  Susceptible ($S$), the  fraction of the total population that is
vulnerable and at a risk of being infected. The Infective ($I$), the population that has been tested positive for infection. The Recovered or Removed ($R$), the
population that has either recovered and reported negative or lost their lives. Here, we assume that a recovered person will not be infected again, because of developing immunity and through informed preventive measures and isolation. The model consists of the following set of non-linear differential equations that can be found e.g. in \cite{Newman}.
\begin{align}
\label{eq:S}\frac{dS}{dt}=&-\beta S(t)I(t)
\\\label{eq:I}\frac{dI}{dt}=&\beta S(t)I(t)-\gamma I(t)
\\\label{eq:R}\frac{dR}{dt}=&\gamma I(t)
\end{align}
where $t$ is the  time  (in day), $\beta(t)$ is the infection rate that means the number of persons contracting infection per unit time. The parameter $\gamma(t)$  is the rate at which infected individuals are either recovered or died and are no longer host to the disease. The COVID-19 outbreak is a contiguous disease that has become pandemic quite
quickly worldwide and, therefore, can be modelled with SIR. The set of equations (1-3) are coupled through the terms  $\beta S(t)I(t)$ (the newly added infected cases) and $\gamma I(t)$ (the newly recovered cases). The three groups of populations which are represented by variables $S(t)$, $I(t)$, $R(t)$ and determined by the above set of equations, will sum up to give the total number of population i.e.
\begin{align}
 \label{eq:N}S(t)+I(t)+R(t)=N.
\end{align}
In the spirit of the model developed in \cite{LZhong} we take the total population of Pakistan as susceptible, however, the lock-down and social  distancing measures reduced the size of susceptible population but  a large number still remains vulnerable. 
Initially the  number of confirmed infected cases is very low, and most of the population are in the suspectable state, therefore the $S(t)$ is assumed as constant equal to the total population $N$ and hence equation~(\ref{eq:S}) can be ignored in the discussion here. In addition, the constant $S$ in equation~(\ref{eq:I}) can be absorbed in  parameter $\beta$. The set of equations in SIR model can be simplified into a discrete version for implementation to the COVID-19 pandemic as the data is provided on daily basis. Equation~(\ref{eq:I}) in the  finite difference form becomes,
\begin{align}\label{eq:Infected}
I(t+\Delta t) = I(t) + \big(\beta-\gamma\big)I(t)\Delta t
\end{align}
where $\Delta t$ is the interval for integration and the change in the infected individuals with unit time depends upon difference of $\beta(t)$ and $\gamma(t)$. From these equations, the expression for $\beta(t)$  is given by, 
\begin{align}\label{eq:beta}
\beta(t) = \frac{[I(t+\Delta t)-I(t)]+[R(t+\Delta t)-R(t)]}{I(t)\Delta t}
\end{align}
The second parameter of the model $\gamma(t)$  is determined from the difference of recoveries given by the expression,
\begin{align}\label{eq:gamma}
\gamma(t) = \frac{R(t+\Delta t)-R(t)}{I(t)\Delta t}
\end{align}
\input{betaGammaTable_singlePage}
%



\begin{figure}[htbp]
\begin{center}
\begin{tabular}{cc}

\def\stackalignment{l}
\topinset{(a)}{
\includegraphics[width=0.47\textwidth, height=0.3\textheight]{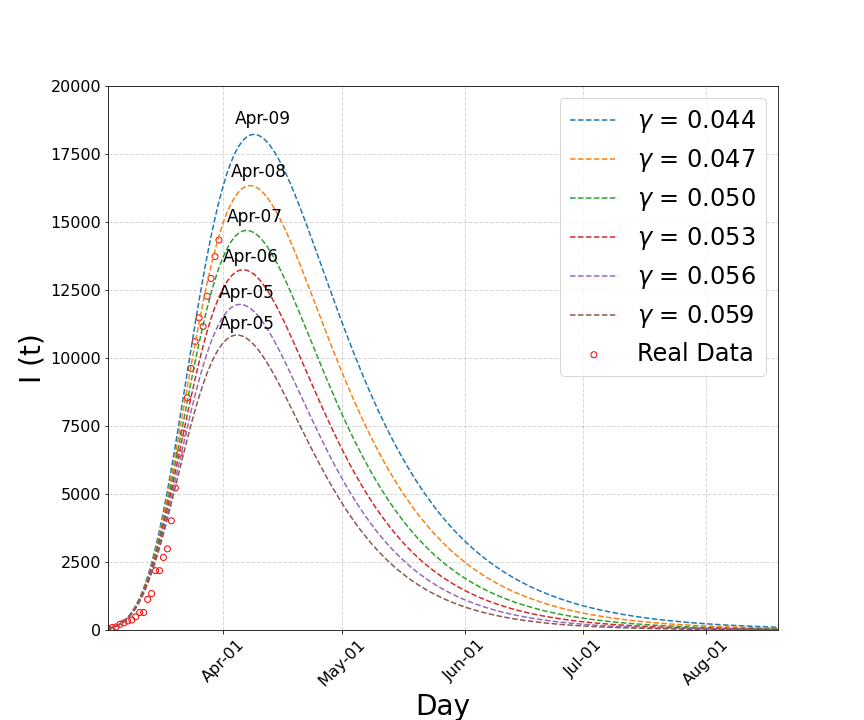}
}{0.5in}{.5in}
\topinset{(b)}{
\includegraphics[width=0.47\textwidth, height=0.3\textheight]{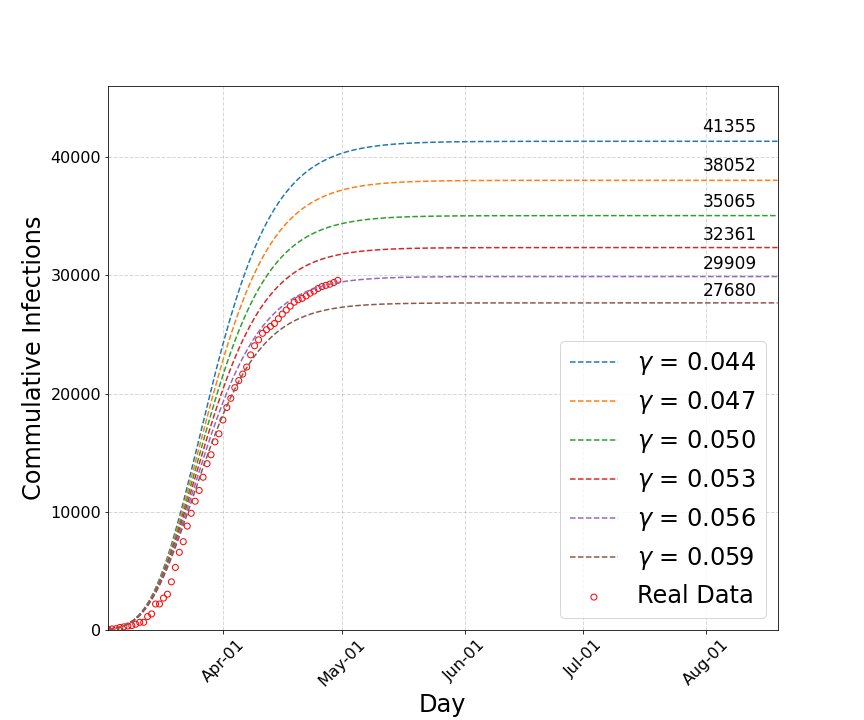}
}{0.5in}{.5in}
\end{tabular}
\end{center}
 \caption{Active and cumulative Infectives as function of time (in day) for Switzerland. (a) shows the active infective cases as function of time. (b) shows cumulative number of infections as a function of time. }
  \label{fig:validat}
\end{figure}

\section{Data Analysis}
\label{sec:dataAna}
Corona Virus Disease that started in early December has become a global pandemic. It has already spread to more than 183 countries in about 3 months time. In Pakistan, the total number of infected cases have increased from zero to more than 19,000 in about two months as shown in Table~\ref{tab:summary}. 
As explained in section~\ref{sec:model}, the time-varying variables namely infection rate $\beta(t)$ and removal rate $\gamma(t)$ are the two very important parameters that have to be estimated from the available epidemiological data. The data used in this paper is taken from the John Hopkins University dashboard~\cite{JH}, which provides comprehensive list of worldwide data in the form of spreadsheet for comparison with other countries and analyses.

One important constraint on the estimation of parameters $\beta(t)$ and $\gamma(t)$ is that these variables should be monotonically decreasing and increasing functions with time, respectively. These constraints ensure the number of Infectives $I(t)$ in the SIR model to follow a bell-shaped curve in accordance with the epidemic transmission model~\cite{KCAng}. On physical ground, the parameter $\beta(t)$ should be monotonically decreasing with time, otherwise, the pandemic will not stop unless all the susceptible population gets infected. The later case will be a very extreme situation, that has not been happening in the epidemics like SARS in 2003 and still not the case in COVID-19 in many countries such as China and Switzerland which are recovering and finishing the disease period.
A second constraint on the parameter  $\beta(t)$ is that it should not be sharply decreasing otherwise the model will significantly underestimate the epidemic transmissibility. A good estimate for the parameter $\beta(t)$ is found by fitting different subsets of data and constraining it by using the above mentioned criterions.

The parameter $\gamma(t)$ is generally very small at the beginning of the epidemic because most Infectives have not recovered or removed yet and hence varies slowly with time. We have tried the time-varying parameterizations of $\gamma(t)$ which did not help significantly. An unreasonably small $\gamma$ values will cause unrealistically long epidemic duration and vice versa.
 In this analysis, we followed a similar strategy for estimating parameters $\beta(t)$ and $\gamma(t)$ from the data as discussed in~\cite{LZhong}. Models are scanned for different fixed values of $\gamma(t)$ and the uncertainty in the prediction of Infectives is presented for a window of removal rates. The interval for $\gamma$ must contain the model which fits reasonably well to the data.

 To predict the peak position of the distribution of Infective cases versus time, we also use another variable called the reproduction number $R_{0}(t)$, which is defined as the ratio of $\beta(t)$ and $\gamma(t)$. The reproduction number is also time-dependent variable. Initially, when the infection rate $\beta(t)$ is large and removal rate $\gamma(t)$ is small, the $R_{0}(t)$ is expected to be large, which decreases gradually and will become zero at the end of the epidemic. The peak of infective cases can be defined as the day when the $R_{0}(t)$ becomes equal to unity, afterward the removal rate dominates and the number of Infectives start decreasing.
\begin{figure}[htbp]
\begin{center}
\begin{tabular}{cc}

\def\stackalignment{l}
\topinset{(a)}{
\includegraphics[width=0.47\textwidth, height=0.3\textheight]{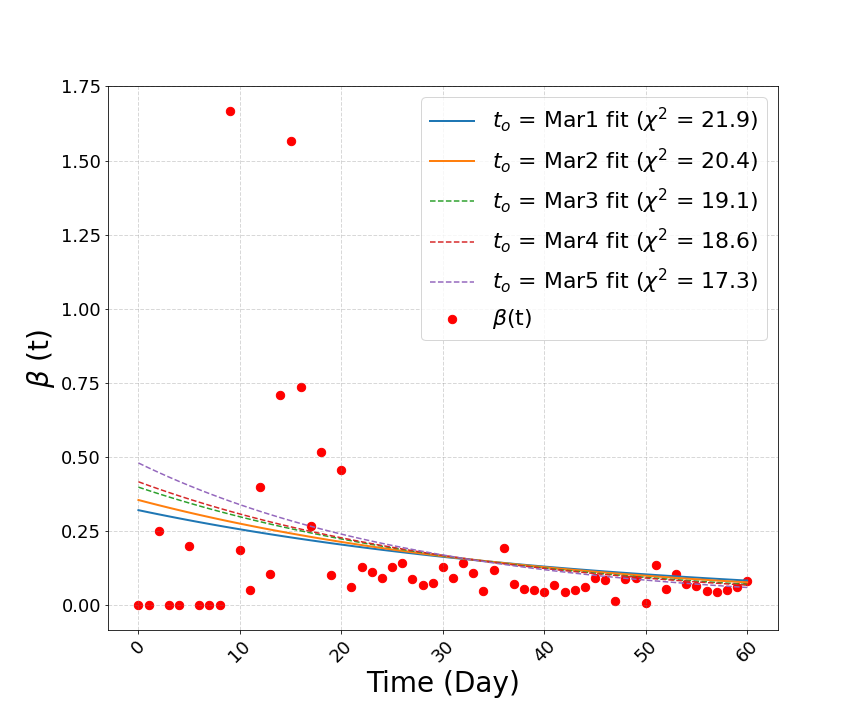}
}{0.5in}{.5in}
\topinset{(b)}{
\includegraphics[width=0.47\textwidth, height=0.3\textheight]{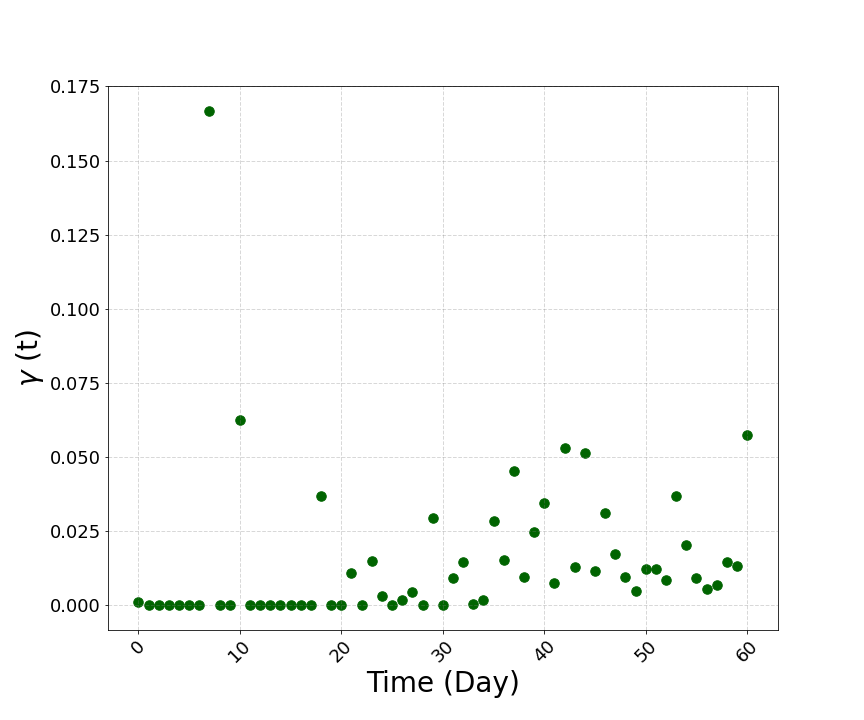}}{0.5in}{.5in}

\end{tabular}
\end{center}
 \caption{Infection and recovery rates as a function of time (in day). (a) shows infection rate ($\beta$) as a function of time. (b) shows recovery rate(recovery+death) as function of time. The lines show exponential fit performed to parameterize $\beta$ as function of time. }
  \label{fig:beta:gamma}
\end{figure}
\begin{figure}[htbp]
\begin{center}
\begin{tabular}{cc}

\def\stackalignment{l}
\topinset{(a)}{
\includegraphics[width=0.47\textwidth, height=0.3\textheight]{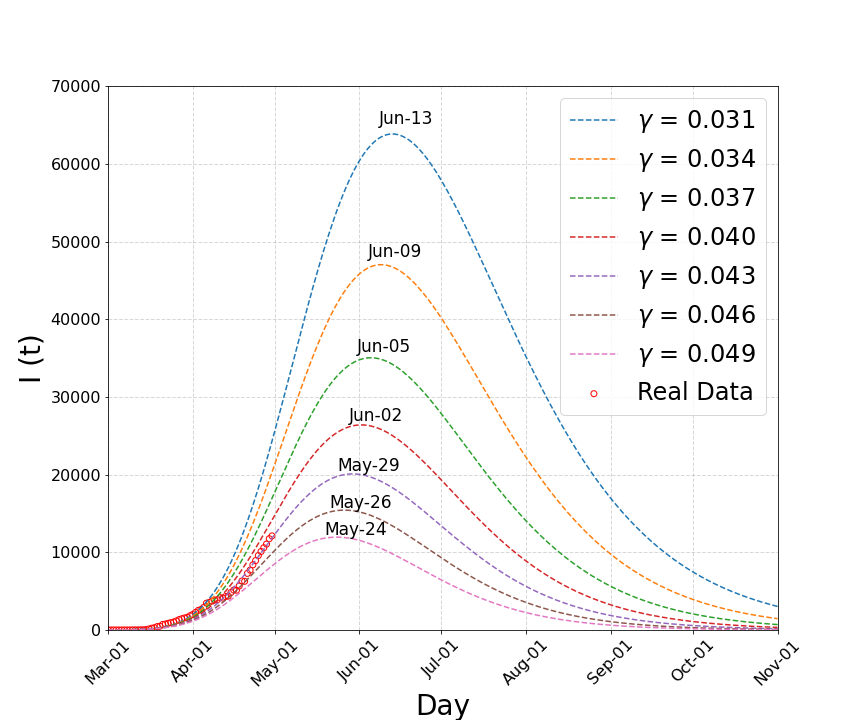}
}{0.5in}{.5in}
\topinset{(b)}{
\includegraphics[width=0.47\textwidth, height=0.3\textheight]{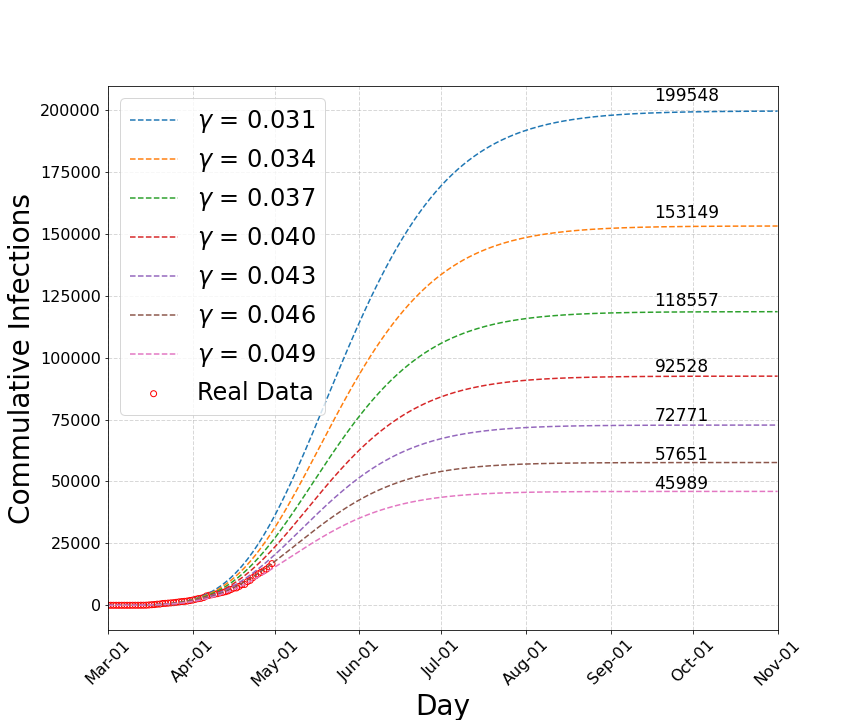}
}{0.5in}{1.2in}
\end{tabular}
\end{center}
 \caption{Active and cumulative Infectives as a function of time (in day) for Pakistan. (a) shows the active infection cases as function of time for different values of recovery rates. (b) shows cumulative number of infections as a function of time for different values of recovery rates. }
  \label{fig:infectVsday}
\end{figure}
\begin{table}[tb]
\begin{center}
\caption{Summary of predictions of COVID-19 pandemic for European Countries. Number of Infectives in data are those recorded on  3 May 2020.}
\label{tab:valid}
\begin{tabular}{ l  c  c | c }
\hline
Country & Peak date Interval & Peak Infectives(Data) & Cumulative Infectives(Data)\\
\hline
\hline
 Switzerland &  7-8 April &   14705-16352  (14349) & 29909-35065 (29905)\\
 Spain & 11-12 April &      79742-90956 (87312) & 236471-295607 (217466)\\
 Italy & 12-14 April &      94666-108980(108257) & 220863-245331(210717)\\
 Germany & 13-14 April &      70711-81058 (72864)& 165047-205423(165664)\\
  \hline
\end{tabular}
\end{center}
\end{table}
\section{Analysis Validation}
\label{sec:anaValid}
To cross-check our analysis, we have applied the same machinery as used in this paper to 
countries where the pandemic period has either ended or in the finishing phase. For the validation purpose of our analysis procedures and tools, we have reproduced some of the predictions for Chinese pandemic (plots not shown) and performed analysis for Switzerland, Italy, Germany and Spain. For the last four countries a summary is given in Table~\ref{tab:valid}. The numbers agree reasonably well with the data for these countries, except the prediction of the peak date of the pandemic which is based on the distribution of active infected cases as shown in Figure~\ref{fig:validat}(a). The peak date of the pandemic is off (overestimated) by a week with respect to the peak position in the data for Switzerland. The reason for discrepancy could be due to the fixed value of $\gamma$  being used in the model instead of the time dependent parametrization. Another method(plots not shown) which has been studied for predicting the peak of the pandemic, is by comparing the number of daily new infections with the SIR model. That distribution gives the peak of the pandemic to fall in the mid May but the method itself is very sensitive to the fluctuations in data. On the other hand the cumulative active infection distribution is not very sensitive to any fluctuations in data(which may arises due to changes in the testing capacity/strategies or sudden spikes in the number of infections) and is therefore used as baseline method in this analysis. To take the discrepancy into account, an uncertainty of one week is assigned to the peak date of the pandemic determined with the default method used in this paper.
\par The distribution of active Infectives and cumulative Infectives as a function of time for Switzerland are shown in Figure~\ref{fig:validat}. There has been a very good agreement between the data points and the SIR model for a recovery rate of 0.047. The inflection point of peak of pandemic occurred around 8 April and the model predicts around 29909 cumulative Infective cases in Switzerland, which can be compared to the total cumulative infection of 29905 on 3rd May with almost flatten curve. The results validate our analysis procedure and give us confidence to accurately predict the pandemic scenarios for Pakistan.

\section{Results and Discussions}
\label{sec:results}
The input epidemiological data which has been used in this paper is shown in Table~\ref{tab:summary}. The cumulative infected cases in Pakistan on March 1st and 2nd were 4 and, therefore the infection rate $\beta(t)$ is zero. On March 5, the cumulative Infective increased to 5 and, therefore, the infection rate $\beta(t)$ jumps to 0.25. For the following two days, the number of Infective remains constant and thus $\beta(t)$ is zero. The recovery rate $\gamma(t)$ remains zero for obvious reasons and for the first time jumps to 0.17 when one person recovers on March 8. The infection rate jumps sharply to 1.67 on March 10 when the number of cumulative Infective becomes almost triple the number of cases on the previous day. During the following days, $\beta(t)$ drops sharply once again. On March 16, there is a spike in the number of Infectives which makes $\beta(t)$ to rise sharply. Such a sudden rise in $\beta(t)$ due to change in the number of Infectives could be attributed to sharp increase in the number of tests performed or testing a large group of people arriving from abroad.
To estimate the time-dependent infection and recovery rates, we pass cumulative Infectives data per day as input to SIR model. The parameters $\beta(t)$ and $\gamma(t)$ are estimated by using equations~(\ref{eq:beta}) and~(\ref{eq:gamma}).  
The values of $\beta(t)$ and $\gamma(t)$ obtained are shown as red circles in Figure~\ref{fig:beta:gamma}(a) and green circles in Figure~\ref{fig:beta:gamma}(b) respectively.

\begin{figure}[htbp]
\begin{center}
\begin{tabular}{cc}

\def\stackalignment{l}
\topinset{(a)}{
\includegraphics[width=0.47\textwidth, height=0.3\textheight]{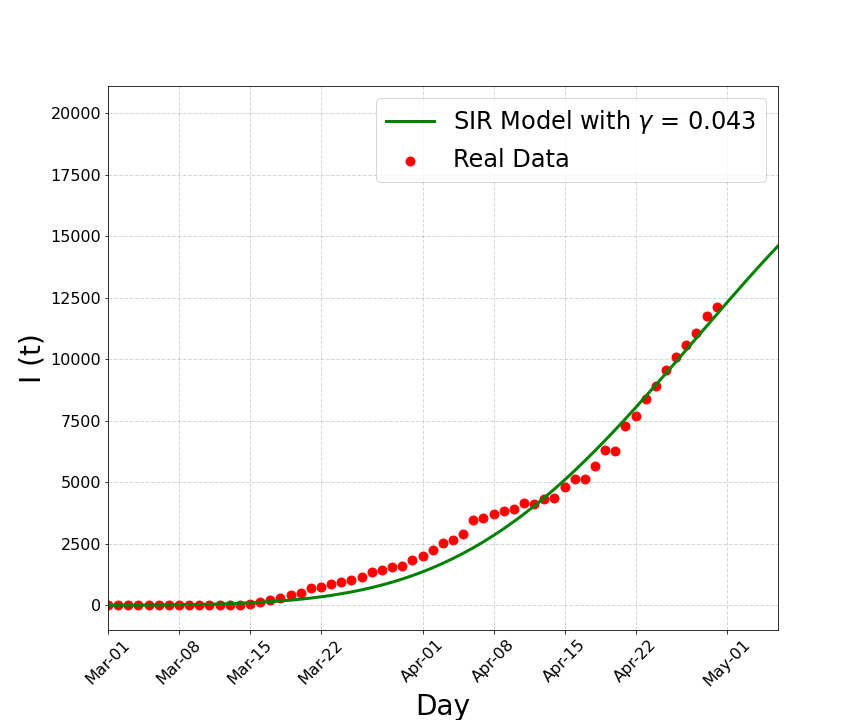}
}{0.5in}{.5in}
\topinset{(b)}{
\includegraphics[width=0.47\textwidth, height=0.3\textheight]{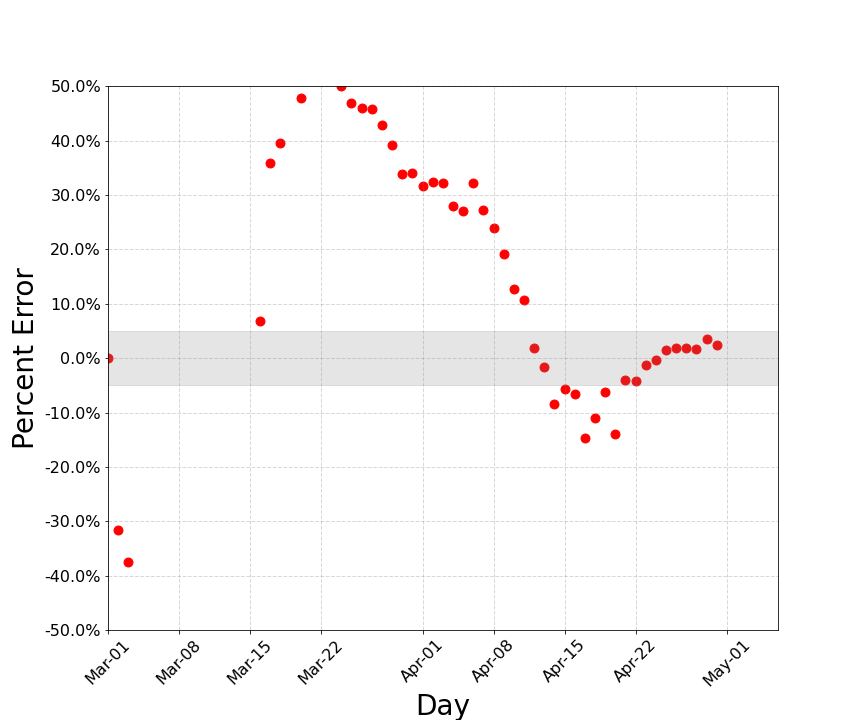}
}{0.5in}{.5in}

\end{tabular}
\end{center}
 \caption{Comparison of predictions and data for active infected cases. (a) shows the active infected people in data and model for $\gamma$ = 0.043. (b) shows the percentage error on prediction. Initially there are large fluctuations but after April 15, the model agrees with data within 5\% uncertainty band. }
  \label{fig:predictErr}
\end{figure}

\begin{figure}[htbp]
\begin{center}
\begin{tabular}{cc}

\includegraphics[width=0.47\textwidth, height=0.3\textheight]{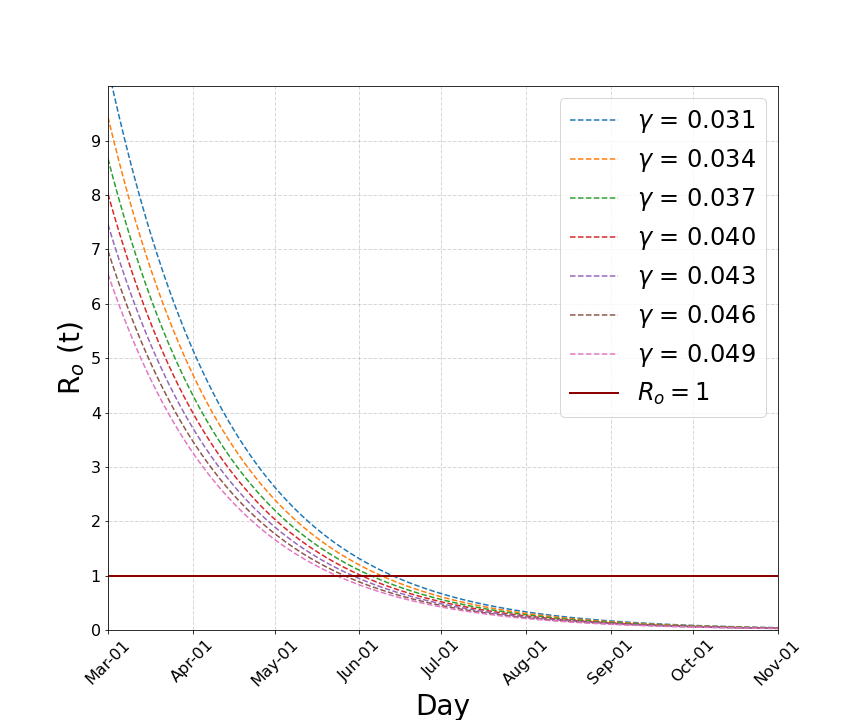}

\end{tabular}
\end{center}
 \caption{Reproduction number vs time (in day). The horizontal solid line shows the turning point when the recovery rate dominates the infection rate. The SIR model for different $\gamma(t)$ values passes through the horizontal line in the time window from 29 May to 9th June, which marks the turning point of the pandemic. }
  \label{fig:R0}
\end{figure}

A time-dependent parametrization of $\beta(t)$ is obtained by performing exponential fits to the infection rate data as shown in Figure~\ref{fig:beta:gamma}(a). To find the best fit, several subsets of data are extracted by sequentially excluding the data between 1st March to 5th March in the one-day interval. The results of exponential fits for five subsets of data are shown in Figure~\ref{fig:beta:gamma}(a). All the fits are having negative slops which give rise to $\beta(t)$ decreasing with time. The fits corresponding to subsets of data from 2nd March to 5th March are excluded due to rapidly falling slope, as they will underestimate the peak and cumulative infected cases, and the ending of the pandemic.  For pandemic predictions and evolution in coming months, parametrization obtained based on the data set of 1st March, is passed to the SIR model as time dependent parametrization of $\beta(t)$. The exponential fit corresponding to 1st March is shown as solid line in Figure~\ref{fig:beta:gamma}(a). The exponential function based on dataset starting from 1st March essentially includes all the data since the start of the pandemic in Pakistan when four infected cases were reported.

The variable $\gamma(t)$ is obtained from data by using equation~(\ref{eq:gamma}) and plotted in Figure~\ref{fig:beta:gamma}(b).
The parameter $\gamma(t)$ is a slow varying variable with average value of around 0.02. In this analysis different fixed values of $\gamma(t)$ are used instead of the time-dependent parameterizations to find an optimum window where accurate predictions can be made. For a range of recovery rates (different values of $\gamma(t)$), a plot of active Infectives, which is defined as the difference of cumulative Infectives minus recovered cases, is shown in Figure~\ref{fig:infectVsday}(a). The active infected cases agree reasonably well with the predicted SIR model distribution for a recovery rate of 0.043. The peak of the pandemic will occur at 29 May with active infected cases of around 20000. The peak Infectives exclude the number of cases who either recovered or died as they no more contribute to the transmission of infection. The modelling has been done based on the current data which corresponds to the strict lock-down scenario. Any change in the lock-down and social distancing scenarios could rapidly escalate the infected cases which will shift the peak of Infectives accordingly. Therefore the prediction has to be made in a window of models for different recovery rates. For recovery rates from 0.043 to 0.034, the peak of the pandemic will occur between 29 May and 9 June, 2020. Comparison of data with predictions and the trend of active infection rate don't support models with $\gamma(t)$ larger than 0.049. Those models will significantly underestimate the cumulative number of Infectives and hence excluded from the discussion here.
\begin{table}[tb]
\begin{center}
\caption{Summary of predictions of COVID-19 pandemic for Pakistan.}
\label{tab:limits}
\begin{tabular}{ l  c  c | c }
\hline
Prediction interval & Peak date & Peak Infectives & Cumulative Infectives \\
\hline
\hline
 Lower bound & 29 May &      20094  ($\gamma$=0.043) & 57651 ($\gamma$=0.046)\\
 Upper bound & 9 June &      47043 ($\gamma$=0.037) & 153149 ($\gamma$=0.034)\\
 \hline
\end{tabular}
\end{center}
\end{table}
The cumulative Infective cases as a function of time are shown in Figure~\ref{fig:infectVsday}(b). The figure shows that the cumulative curves will start flattening in June-July. For a change of 0.003 in the recovery rate there is a significant change in the total number of cumulative Infectives as well as flattening of the curves. This suggests that better health care and recovery rates can significantly reduce the total number of infected people and can shorten the duration of the pandemic. The total number of infected cases could be 57651 for a recovery rate of 0.046 and can rise to 153149 for a recovery rate of 0.034. The total number of Infectives can easily cross a figure of 100,000 if the recovery gets worsen. The cumulative infections for upper and lower bounds are summarized in Table~\ref{tab:limits}. A comparison of the data and model prediction is zoomed-in in Figure~\ref{fig:predictErr}(a). The disagreement between 15 March and 10 April could be attributed to the change in health care strategies such as improving patients testing capabilities. Through enforcing rigorous  countrywide lock-down, the active infection rate stabilizes after 10 April. The percentage error on the number of Infective per day is shown in Figure~\ref{fig:predictErr}(b). Apart from some outliers, the error on one-day prediction lies within 5\% uncertainty band, which is shown as shaded region in the figure.
To cross-check the turning point or peak position of the model, a variable known as reproduction number is defined as the ratio of infection rate to recovery rate. The plot of reproduction number R$_{0}$ as a function of time is shown in Figure~\ref{fig:R0}. The solid horizontal line shows the turning point when the recovery rate will dominate the infection rate. At this point, the pandemic will turn around as more people will recover than infected and this also marks the peak position in the SIR model. For different recovery rates the models cross the horizontal line in the time window from late May to 9th June, which marks the turning point of the pandemic. The reproduction number R$_{0}$ becomes less than unity after the peak or turning point of the outbreak and the pandemic is expected to fade away with 97\% of the recovery already happened by August-September 2020.


\section{Conclusion}
\label{sec:conclusion}
The modelling and prediction of COVID-19 for Pakistan have been presented by using a simple time dependent SIR model. The two input parameters namely the infection rate $\beta(t)$ and recovery rate $\gamma(t)$ are estimated from data using  equations~(\ref{eq:beta}) and~(\ref{eq:gamma}). A time-dependent parametrization of $\beta(t)$ is obtained by performing exponential fit to the data. The analysis has been validated by predicting the pandemic peak and cumulative Infectives of Switzerland and other European countries. The cumulative Infectives for Switzerland are predicted to be around 29909 when the curves will flatten in the first half of May. This prediction is in good agreement with the total Infectives of 29905 on 3rd May.  The analysis is then performed to predict the pandemic scenario in Pakistan. A comparison of active infection rate for data and models are made for different values of recovery rate $\gamma(t)$. The best match occurs for the model with a recovery rate of 0.043 for unrecovered peak Infectives of 20094, which is  expected to happen on 29 May(with uncertainty of a week on the peak date). Based on a window of recovery rate $\gamma(t)$, the prediction interval of turning or peak point of the pandemic is expected to occur from late May to 9 June with unrecovered number of Infectives from 20000-47000. The cumulative number of Infectives in the existing lock-down scenario is predicted to be in the range 57651-153100. These numbers are very sensitive to both transmission and recovery rates. For example, a 1\% reduction in the recovery rate will triple the number of cumulative infected people. The number of Infectives will remain at the lower end under the strict preventive scenario and can rapidly double or triple if the spread of the epidemic is not contained through strict measures. To keep the cumulative Infectives around 60000 and to get rid of the pandemic by August-September, it is proposed to keep strict preventive measures intact till mid-June, 2020.

The turning point of the pandemic is also modelled with the reproduction number  R$_{0}$ which predicts the same time window as given by the SIR model. The reproduction number R$_{0}$ becomes less than unity after the peak of the active Infective distribution. The pandemic is expected to fade away completely with 97\% Infectives recovered by August-September.


\end{document}

%% file: betaGammaTable_singlePage.tex

\begin{table}
\begin{center}
\caption {\small{Epidemiological data of the COVID-19 pandemic, cumulative infected rate is represented by I(t), cumulative recovery rate including both dead and recovered cases are denoted as R(t).}  }
\label{tab:summary}
\begin{tabular}{|l l l l l |l l l l l|}
  \hline
 \textbf{Date} &\textbf{I(t)}  & \textbf{R(t)} & \textbf{$\beta$(t)} & \textbf{$\gamma$(t)} & \textbf{Date} &\textbf{I(t)}  & \textbf{R(t)} & \textbf{$\beta$(t)} & \textbf{$\gamma$(t)}  \\
 \hline
3/1/20 &4 &0(0+0)&0.00&0.00&4/1/20&2118&121(94+27)&0.09&0.00\\
3/2/20 & 4 & 0(0+0)&0.00&0.00&4/2/20&2421&159(125+34)&0.14&0.01\\
3/3/20 &5 &0(0+0)&0.25&0.00&4/3/20&2686&166(126+40)&0.11&0.00\\
3/4/20 & 5&0(0+0)&0.00&0.00&4/4/20&2818&172(131+41)&0.05&0.00\\
3/5/20&5&0(0+0)&0.00&0.00&4/5/20&3157&258(211+47)&0.12&0.03\\
3/6/20&6&0(0+0)&0.20&0.00&4/6/20&3766&312(259+53)&0.19&0.02\\
3/7/20 & 6&0(0+0)&0.00&0.00&4/7/20&4035&486(429+57)&0.07&0.05\\
3/8/20&6&1(1+0)&0.00&0.17&4/8/20&4263&528(467+61)&0.06&0.01\\
3/9/20&6&1(1+0)&0.00&0.00&4/9/20&4489&637(572+65)&0.05&0.02\\
3/10/20&16&1(1+0)&1.67&0.00&4/10/20&4695&793(727+66)&0.05&0.03\\
3/11/20&19&2(2+0)&0.19&0.06&4/11/20&5011&848(762+86)&0.07&0.01\\
3/12/20&20&2(2+0)&0.05&0.00&4/12/20&5230&1119(1028+91)&0.04&0.05\\
3/13/20&28&2(2+0)&0.40&0.00&4/13/20&5496&1188(1095+93)&0.05&0.01\\
3/14/20&31&2(2+0)&0.11&0.00&4/14/20&5837&1474(1378+96)&0.06&0.05\\
3/15/20&53&2(2+0)&0.71&0.00&4/15/20&6383&1557(1446+111)&0.09&0.01\\
3/16/20&136&2(2+0)&1.57&0.00&4/16/20&6919&1773(1645+128)&0.08&0.03\\
3/17/20&236&2(2+0)&0.74&0.00&4/17/20&7025&1900(1765+135)&0.02&0.02\\
3/18/20&299&2(2+0)&0.27&0.00&4/18/20&7638&1975(1832+143)&0.09&0.01\\
3/19/20&454&15(13+2)&0.52&0.04&4/19/20&8348&2036(1868+168)&0.09&0.00\\
3/20/20&501&16(13+3)&0.10&0.00&4/20/20&8418&2146(1970+176)&0.01&0.01\\
3/21/20&730&16(13+3)&0.46&0.00&4/21/20&9565&2274(2073+201)&0.14&0.01\\
3/22/20&776&10(5+5)&0.06&0.01&4/22/20&10076&2357(2156+201)&0.05&0.01\\
3/23/20&875&11(5+6)&0.13&0.00&4/23/20&11155&2764(2527+237)&0.11&0.04\\
3/24/20&972&25(18+7)&0.11&0.01&4/24/20&11940&3008(2755+253)&0.07&0.02\\
3/25/20&1063&29(21+8)&0.09&0.00&4/25/20&12723&3135(2866+269)&0.07&0.01\\
3/26/20&1201&30(21+9)&0.13&0.00&4/26/20&13328&3217(2936+281)&0.05&0.01\\
3/27/20&1373&34(23+11)&0.14&0.00&4/27/20&13915&3321(3029+292)&0.04&0.01\\
3/28/20&1495&41(29+12)&0.09&0.00&4/28/20&14612&3545(3233+312)&0.05&0.01\\
3/29/20&1597&43(29+14)&0.07&0.02&4/29/20&15525&3768(3425+343)&0.06&0.01\\
3/30/20&1717&97(76+21)&0.08&0.03&4/30/20&16817&4658(4315+343)&0.08&0.06\\
3/31/20&1938&102(76+26)&0.13&0.00& & & & &  \\
\hline
\end{tabular}
\end{center}
 \end{table}


%% file: COVID19_Analysis_Pakistan_SIR.bbl
\begin{thebibliography}{1}
\bibitem{Lina} Q. Lina, S. Zhaob, D. Gaod, Y. Loue, S. Yangf, S. S. Musae, M. H. Wangb, Y. Caig, W. Wangg, L. Yangh, D. Hee, A conceptual model for the coronavirus disease 2019 (COVID-19) outbreak in Wuhan, China with individual reaction and governmental action.
\href{https://www.ijidonline.com/article/S1201-9712(20)30117-X/fulltext}{International Journal of Infectious Diseases, vol. 93, pp. 211-216,  2020}.

\bibitem{LZhong} L. Zhong, L. Mu, J. Li, J. Wang, Z. Yin, and D. Liu, Early Prediction of the 2019 Novel Coronavirus Outbreak in the Mainland China Based on Simple Mathematical Model.
    \href{https://ieeexplore.ieee.org/document/9028194}{in IEEE Access, vol. 8, pp. 51761-51769, 2020}.

\bibitem{YChen} Yi-Cheng Chen, Ping-En Lu, Cheng-Shang Chang, and Tzu-Hsuan Liu, A Time-dependent SIR model for COVID-19 with Undetectable Infected Persons.
   \href{http://gibbs1.ee.nthu.edu.tw/A\ TIME\ DEPENDENT\ SIR\_MODEL\_FOR\_COVID\_19.PDF}{Link to the latest version of paper}

\bibitem{Newman} M. Newman, Networks: An Introduction. \href{https://www.oxfordscholarship.com/view/10.1093/acprof:oso/9780199206650.001.0001/acprof-9780199206650}{Oxford University Press, 2010}.

\bibitem{JH} CSSEGISandData and J. H. University, “Covid-19,” Feb 2020. [Online]. Available: \href{https://github.com/CSSEGISandData/COVID-19}{https://github.com/CSSEGISandData/COVID-19} .
\bibitem{SIRModel} X. N. Han, S. J. De Vlas, L. Q. Fang, D. Feng, W. C. Cao, and J. D. F. Habbema, ``Mathematical modelling of SARS and other infectious diseases in China: A review,'' vol. 14, no. s1, pp. 92-100, 2009, doi:10.1111/j.1365-3156.2009.02244.x.


\bibitem{KCAng} K. C. Ang, ``A simple stocchastic model for an epidemic-numerical experiments with MATLAB,'' Electron. J. Math. Technol., vol. 1, no. 2,pp. 117-128, 2007.

\end{thebibliography}
